# Scene-and-Process-Dependent Spatial Image Quality Metrics

Edward W. S. Fry*[1], Sophie Triantaphillidou[1], Robin B. Jenkin[2], Ralph E. Jacobson[1], John, R. Jarvis[1]

[1]Computational Vision and Imaging Technology Research Group, University of Westminster, London, UK
[2]NVIDIA Corporation, Santa Clara, CA, USA
*Corresponding author (Email: e.fry@my.westminster.ac.uk)

## ABSTRACT

Spatial image quality metrics designed for camera systems generally employ the Modulation Transfer Function (MTF), the Noise Power Spectrum (NPS), and a visual contrast detection model. Prior art indicates that scene-dependent characteristics of non-linear, content-aware image processing are unaccounted for by MTFs and NPSs measured using traditional methods. We present two novel metrics: the log Noise Equivalent Quanta (log NEQ) and Visual log NEQ. They both employ scene-and-process-dependent MTF (SPD-MTF) and NPS (SPD-NPS) measures, which account for signal-transfer and noise scene-dependency, respectively. We also investigate implementing contrast detection and discrimination models that account for scene-dependent visual masking. Also, three leading camera metrics are revised that use the above scene-dependent measures. All metrics are validated by examining correlations with the perceived quality of images produced by simulated camera pipelines. Metric accuracy improved consistently when the SPD-MTFs and SPD-NPSs were implemented. The novel metrics outperformed existing metrics of the same genre.

*Keywords:* Image Quality, Image Quality Metric (IQM), Modulation Transfer Function (MTF), Noise Power Spectrum (NPS), Noise Equivalent Quanta (NEQ), Contrast Sensitivity Function (CSF), Scene-Dependency, Log Noise Equivalent Quanta Metric (Log-NEQ), Scene-and-Process-Dependent MTF (SPD-MTF), Scene-and-Process-Dependent NPS (SPD-NPS),

## Introduction

Subjective image quality is defined by Engeldrum as "the integrated set of perceptions of the overall degree of excellence of an image" [1]. It is often expressed as the multivariate combination of visuo-cognitive factors – the "nesses" – concerning image quality attributes such as resolution, sharpness, noisiness, contrast and colorfulness [1]. Spatial image quality relates to the intensity and distribution of two-dimensional (2D) luminance contrast signals and is associated with sharpness, resolution, noise and contrast.

Psychophysical evaluations are the only true means of measuring the overall subjective quality of images, or the perceived magnitude of their individual quality attributes. Different methods involve categorical scaling [1], paired/triplet comparison [2] or comparison with a calibrated series of ruler images using the ISO 20462 [3] Image Quality Ruler. The last records observer ratings on a generalizable ratio Subjective Quality Scale (SQS2) [3], with increments of just-noticeable-difference (JND), and a zero point that refers to low-quality scenes that are difficult to identify perceptually.

Despite recent developments in speed and precision [2], [3], psychophysical image quality evaluations are slow, expensive and difficult to carry out accurately. IQMs save this time and expense by mapping the image information – or data concerning imaging system performance and the viewing conditions – to output scores that are intended to correlate with observer quality ratings, Figure 1.

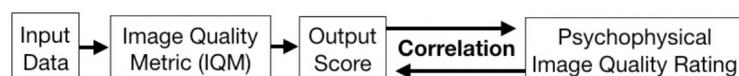

*Figure 1. Generalization of image quality metric (IQM) characteristics [4].*

IQMs are employed for online image quality control, image signal processing (ISP) algorithm development and imaging systems design and optimization. A broad spectrum of metrics has evolved from several research areas including imaging systems engineering, information theory, signal/image processing, computer vision, machine learning, visual psychophysics and neural physiology [5]. These IQMs are tailored for different applications and apply varying levels of calibration, or curve-fitting, to optimize their correlation with observer quality ratings from test image datasets that contain different types of artefacts.

This paper is concerned specifically with no-reference spatial metrics suited for image capture systems engineering. Suitable IQMs break image quality judgement down into components relating to the different attributes, and the characteristics of imaging system components and the human visual system (HVS). A recent review [4] by the authors defines the following spatial IQM genres: *Computational IQMs*, *Image Fidelity Metrics, Signal Transfer Visual IQMs (STV-IQM),* and *Multivariate Formalism (MF-IQM)*. When each genre was evaluated from a capture system engineering perspective, the Computational IQMs and Image Fidelity Metrics were concluded to be least suitable for the purpose [4].

The STV-IQMs and MF-IQMs – referred to in this paper as *engineering metrics* – employ standard spatial system performance measures such as the Modulation Transfer Function (MTF) and Noise Power Spectrum (NPS), and threshold contrast sensitivity functions (CSF) describing visual spatial sensitivity. The Noise Equivalent Quanta (NEQ) signal-to-noise measure is core to the most relevant STV-IQMs and is applied widely in capture system and sensor modelling [6]–[8]; it also uses the MTF and NPS.

Our recent evaluation of simulated camera pipelines, however, revealed that the currently employed MTF and NPS measures characterize systems using non-linear content-aware image signal processing (ISP) with limited accuracy, and that novel *Scene-and-Process-Dependent MTF (SPD-MTF)* and *NPS (SPD-NPS)* measures are more suitable [9]. Likewise, contextual contrast detection [10] and discrimination [11] models, which account for each scene's contrast spectrum, should be more suitable visual models for image quality analysis than the currently used CSFs.

This paper aims to revise current STV-IQMs and MF-IQMs to use these state-of-the-art input parameters and then validate them against their original incarnations. We also validate two novel STV-IQMs developed in this laboratory – the *log NEQ* and *Visual log NEQ* – that are based upon a similarly revised *Scene-and-Process-Dependent NEQ (SPD-NEQ)* measure.

The following sections of the paper introduce the STV-IQMs and MF-IQMs and discuss the limitations of their current input parameters. We then define the revised NPS, MTF and CSF parameters. The revised SQRIn, PIC and CPIQ metrics, and the novel Log NEQ and Visual Log NEQ metrics that implement these parameters are then defined. The validation methodology is then presented, which involves the comparison of each IQM's score with observer image quality scores for a number of test images generated by two camera simulation pipelines that apply either linear or non-linear ISPs. Finally, we benchmark each IQM's accuracy and further analyze correlation plots describing typical metric behavior.

**Engineering Metrics**

STV-IQMs and MF-IQMs express image quality as a function of the input signal, the system's performance, and the observer's visual sensitivity under the viewing conditions. Their output scores are causally justified, relating to the imaging system and the HVS.

Univariate STV-IQMs, such as Acutance [12] (see later on, Eq. 10) model perceived image quality concerning the sharpness attribute. Multivariate STV-IQMs are of more relevance to this paper. They account for both sharpness and noisiness, building upon signal-to-noise relationships from communications theory [13] and the founding work of Schade [14] and Nelson [15]. In this paper, we revise the following STV-IQMs: Barten's *[16] Square Root Integral with Noise (SQRIn)* and Töpfer and Jacobson's [17] *Perceived Information Capacity (PIC).* Their advantage is simplicity since they apply limited calibration and relate closely to the NEQ [18].

Keelan's MF-IQM [19] predicts the overall quality loss as the Minkowski combination [20] of several perceptually calibrated quality loss metrics for different attributes/artefacts. The recent IEEE P1858 Camera Phone Image Quality (CPIQ) standard [12] defines individual attribute metrics for texture loss, edge Spatial Frequency Response (SFR), local geometric distortion, visual noise, color uniformity, chroma level and lateral chromatic displacement. These metrics were combined to predict overall quality, validated in reference [21]. The CPIQ overall quality loss metric used in this paper – referred to as *the CPIQ metric* – employs the texture loss and visual noise attribute metrics only, thus modelling the perceived quality concerning sharpness and noisiness. It is more complicated to implement and computationally intensive than multivariate STV-IQMs and also applies higher levels of calibration (curve-fitting).

*Engineering Metric Input Parameters*

The accuracy of the engineering metrics is dependent on the capability of their MTF, NPS and CSF parameters to describe the system's performance in terms of sharpness and noise as well as the observer's visual sensitivity, respectively. In this section, we present sources of inaccuracy in the MTF and NPS measures currently employed. Such inaccuracies mainly relate to the MTF and NPS being used to characterize non-linear, content-aware systems, when both measures originate from linear system theory, which requires systems to be linear, spatially invariant and homogenous [18]. We also discuss the theoretical limitations of currently used CSFs.

The discrete 2D NPS, Eq. 1, characterizes the power of the system's noise, $NPS(u,v)$, versus spatial frequency, $(u,v)$, using the discrete Fourier transform (DFT) [22]. $I(x,y)$ is a luminance noise image of size $M \times N$, given by Eq. 2. $g(x,y)$ is the output image intensity and $\bar{g}(x,y)$ the expected intensity. If $I(x,y)$ is a scene, $NPS(u,v)$ is its DFT power spectrum, or $PS(u,v)$. The rotational average of $NPS(u,v)$ or $PS(u,v)$ yields the 1D NPS, $NPS(u)$, or 1D scene power spectrum, $PS(u)$, respectively.

$$NPS(u,v) = \left| \sum_{x=\frac{M}{2}+1}^{M/2} \sum_{y=\frac{N}{2}+1}^{N/2} I(x,y) e^{-2\pi i (ux+vy)} \right|^2 \quad (1)$$

$$I(x,y) = g(x,y) - \bar{g}(x,y) \quad (2)$$

The STV-IQMs currently employ NPSs derived from captured uniform luminance patches, meaning $\bar{g}(x,y)$ can be conveniently assumed constant under certain conditions. The CPIQ visual noise attribute metric is also derived from captured uniform patches [12]. For all capture systems, however, the amount of noise introduced to uniform patches and real scenes is not necessarily the same, since photon noise is a function of intensity. Further inaccuracies result from the application of non-linear content-aware denoising and sharpening ISPs, which modify the intensity and spatial distribution of the noise dependent upon the local image structure [9], rendering it both local-content-dependent, and scene-dependent. Uniform patches provide, in theory, the ideal input signal for content-aware denoising algorithms, and the derived noise image [4] or NPS [9] underestimate the average real-world noise level of such systems. More recently, noise measures [23] have been derived using the more suitable dead leaves target that simulates the "average scene" power spectrum and NSS [24]; but they have not been benchmarked.

The MTF – and the comparable SFR – characterize the system's signal transfer capabilities versus spatial frequency. The STV-IQMs currently employ MTFs derived from either: i) the system's line spread function, measured from a captured "perfect" edge [25]. ii) comparisons of output-to-input modulation of captured sinusoids [18]. Or iii) output-to-input comparisons of (white) noise power spectra [26]. MTFs derived from such signals are not suitable for characterizing the average real-world performance of scene-dependent systems that apply non-linear content-aware ISPs. For example, their failure to account for the JPEG algorithm's non-linearity was found to reduce PIC and SQRIn metric accuracy when modelling the perceived quality of compressed images [26], [27].

Various MTF implementations employ the dead leaves target to attempt to trigger content-aware ISPs at similar levels to the "average natural scene", thus deriving a more suitable average real-world MTF. The intrinsic dead leaves implementation compares output-to-input dead leaves cross spectra [28]. It is tolerant to noise but is a less suitable input parameter for IQMs, since reversible ISPs are often unaccounted for, e.g. contrast stretching or sharpening [29]. The direct dead leaves implementation [30], Eq. 3, is used by the CPIQ [12] texture loss attribute metric. In Eq. 3, $PS_{Input}(u)$ and $PS_{Output}(u)$ are the input and output 1D dead leaves power spectra, respectively, and $u$ is the spatial frequency. $NPS(u)$ is the uniform patch NPS (which is inaccurate for the reasons stated above).

$$MTF(u) = \sqrt{\frac{PS_{Output}(u) - NPS(u)}{PS_{Input}(u)}} \quad (3)$$

Our recent camera pipeline simulations suggest dead leaves signals may not trigger content-aware ISPs in the same fashion as the "average natural scene" [9]. Further, for scene-dependent systems, measurements from a single test target cannot accurately characterize the system's performance for any given input scene.

Finally, engineering metrics [12], [16], [17] generally employ the CSF that models thresholds of detection of unmasked narrow-band stimuli [31], [32] as a function of spatial frequency. Johnson and Fairchild's Luminance CSF [31] is used by the CPIQ metric. It adapts Movshon's [33] black-box model and accounts for the stimulus' spatial frequency only. The SQRIn and PIC metrics both use Barten's [32] luminance CSF that accounts for the stimulus's spatial frequency, angular size and luminance. The IQMs use the CSF as a frequency domain weighting function. However, CSFs are not transfer functions and do not account for higher-level cognitive processes concerning image quality judgement [34]. Although other black-box visual models often take into account psychological functions [4], there is a lack of alternative mechanistic visual models that account for the psychophysical data and viewing conditions.

The debate is ongoing regarding whether image quality perception involves threshold or suprathreshold visual processes [11], or a combination of both [35]. What can be inferred, however, is that in image quality judgments, the detection and discrimination of visual signals, image attributes and artefacts are primarily contextual processes, which are affected by masking from other image contents, as well as noise.

**Revised Metric Input Parameters**

We define briefly here the various SPD-NPSs and SPD-MTFs [9] that were developed to characterize capture systems utilizing content-aware ISPs (e.g. camera phones and autonomous vehicles). We also discuss the contextual CSF (cCSF) [10] and Visual Perception Function (cVPF) models [11] that account for visual masking. In theory, these parameters account for scene-dependency in spatial imaging system performance and the HVS's sensitivity, something that current engineering IQM parameters do not do.

*Scene-and-Process-Dependent Noise Power Spectra (SPD-NPS)*

The SPD-NPS framework [4], derives the 1D NPS, Eq. 1, of temporally varying noise from a scene-and-process-dependent noise image, $I(x,y)$, that accounts for the effect of content-aware ISPs. Equation 2 computes this noise image when $g(x,y)$ is the output image, resulting from any input signal, and $\bar{g}(x,y)$ is the mean image of several captured replicates. All other parameters are as previously defined. Fixed Pattern Noise (FPN) is unaccounted for.

Three SPD-NPS variants are used in this paper, defined below.

i) The *Dead Leaves SPD-NPS* [9] implements the SPD-NPS framework using the dead leaves target to estimate the average real-world system performance. Our simulations [9] show it is a better noise measure than the uniform patch NPS for non-linear systems.

ii) The *Pictorial Image SPD-NPS* [9] executes the SPD-NPS framework using a pictorial scene as the test image. Our simulations suggest it accounts extensively for scene-dependent variations in temporally varying system noise without suffering from significant bias, provided that 10 or more replicates are employed [9].

iii) The *Mean Pictorial Image SPD-NPS* [9] measures the average real-world performance of the system with respect to a number of pictorial image signals. It is computed as the mean of the pictorial image SPD-NPSs over a large image set representing commonly captured images. It accounts for scene-dependency more comprehensively than the dead leaves SPD-NPS.

**Table 1. Summary of Noise Measure Parameters**

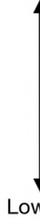

| NPS measure | Summary of characterisation | Accounts for system noise scene-dependency | Sensitivity to system noise scene-dependency |
|---|---|---|---|
| Pictorial Image SPD-NPS | Characterises system noise with respect to a specific image | Yes | Highest |
| Mean Pictorial Image SPD-NPS | Characterises general system noise | Yes | |
| Dead Leaves SPD-NPS | Characterises general system noise | Partially | |
| Uniform Patch NPS | Characterises general system noise | No | Lowest |

*Scene-and-Process-Dependent Modulation Transfer Functions (SPD-MTF)*

The SPD-MTF framework [4] is based upon the direct dead leaves implementation, Eq. 3. $PS_{Input}(u)$ and $PS_{Output}(u)$ are the input and output 1D pictorial scene (or dead leaves) power spectra, respectively. $NPS(u)$ is the pictorial image (or dead leaves) SPD-NPS. All pictorial images are windowed (i.e. their edges are faded to a neutral pixel value) to mitigate bias due to periodic replication artefacts resulting from DFT processing [9].

This paper employs the following three SPD-MTF measures.

i) The *Dead Leaves SPD-MTF* [9] implements the SPD-MTF framework using dead leaves signals to estimate the system's average real-world signal transfer. It uses the dead leaves SPD-NPS and is, therefore, a more appropriate measure for non-linear systems than the direct dead leaves MTF, which uses the uniform patch NPS.

ii) The *Pictorial Image SPD-MTF* [9] executes the SPD-MTF framework using a single pictorial scene and employs the pictorial image SPD-NPS. It characterizes system signal transfer with respect to a given scene, accounting for the scene-dependent characteristics of non-linear ISPs. It builds upon the method of Branca et al. [36] but mitigates significantly biases from periodic replication artefacts and noise underestimation [9]. However, it still suffers from bias due to signal-to-noise limitations that mainly affect higher frequencies of lower-power scenes at lower signal-to-noise ratios (SNR) [9].

iii) The *Mean Pictorial Image SPD-MTF* [9] is computed as the mean of the pictorial image SPD-MTFs across a broad set of typical scenes. Thus, it characterizes the system's average real-world signal transfer, accounting for system scene-dependency. However, the signal-to-noise bias in ii) also occurs in this measure.

Table 2. Summary of Resolution Measure Parameters

| MTF measure | Summary of characterisation | Accounts for system signal transfer scene-dependency? | Accounts for system noise scene-dependency? | Sensitivity to system scene-dependency |
|---|---|---|---|---|
| Pictorial Image SPD-MTF | Characterises system signal transfer for an individual image | Yes | Yes | Highest ↑ |
| Mean Pictorial Image SPD-MTF | Characterises general system signal transfer | Yes | Yes | |
| Dead Leaves SPD-MTF | Characterises general system signal transfer | Partially | Partially | ↓ |
| Direct Dead Leaves MTF | Characterises general system signal transfer | Partially | No | Lowest |

***Contextual Contrast Sensitivity (cCSF) and Visual Perception Functions (cVPF)***

The cCSF [10] and cVPF [11] are based upon Barten's contrast detection [32] and discrimination [35] models, respectively, and employ the Linear Amplification Model [37] to account for visual masking. These scene-dependent functions were validated against observer contrast detection/discrimination datasets, measured from band-limited images of pictorial scenes.

Table 3. Summary of Visual Model Parameters

| Contrast Sensitivity Function (CSF) | Scene-dependent Visual Behaviour Accounted For | Sensitivity to Visual Scene-Dependency |
|---|---|---|
| Johnson & Fairchild CSF | • N/A | Lowest |
| Barten CSF | • Adaptations with respect to global image luminance | ↓ |
| Contextual CSF (cCSF) / Contextual Visual Perception Function (cVPF) | • Adaptations with respect to global image luminance<br>• Contrast masking | Highest |

# Revised Image Quality Metrics

## Scene-and-Process-Dependent SQRIn and PIC

We define the revised SQRIn and PIC metrics, by substituting their current MTF, NPS and CSF parameters with parameters that account for scene-dependent imaging system and/or HVS behavior.

Töpfer and Jacobson's PIC [17], and their reformulation of Barten's [16] SQRIn are defined by Eq. 4 and Eq. 5, respectively. $NPS_{visual}(u)$ is the internal noise power of the eye [38], $CSF(u)$ is the Barten CSF [32], $u$ and $u_{max}$ are the spatial frequency and cut-off frequency, respectively. $k_1$ and $k_2$ are calibration constants. The original definitions of the displayed image spectrum, $S(u)$, and total imaging system noise, $N(u)$, are for analog systems [16], [17].

$$PIC = k_1 \sqrt{\int_0^\infty \ln\left(1 + \frac{S(u)CSF^2(u)}{N(u)CSF^2(u)+NPS_{visual}(u)}\right) \frac{du}{u}} + k_2 \qquad (4)$$

$$SQRIn = \frac{k_1}{\ln 2} \int_0^{u_{max}} \left[\frac{S(u)CSF^2(u)}{N(u)CSF^2(u)+NPS_{visual}(u)}\right]^{0.25} \frac{du}{u} + k_2 \qquad (5)$$

We compute the revised SQRIn and PIC using Eq. 4 and Eq. 5, where $CSF(u)$ refers to either the Barten CSF, cCSF, or cVPF. Revised $S(u)$ and $N(u)$ parameters are used (Eq. 6 and Eq. 8), where $MTF_{SPD}(u)$ and $NPS_{SPD}(u)$ denote the SPD-MTF and SPD-NPS measures, respectively. $PS_{scene}(u)$ is the input scene 1D power spectrum. $\gamma_{disp}$ and $MTF_{disp}(u)$ are the display's gamma and its modelled MTF [12], respectively. $NPS_{disp}(u)$ is the display's NPS; it is assumed negligible in this implementation.

$$S(u) = PS_{scene}(u).MTF_{SPD}^2(u).\gamma_{disp}^2.MTF_{disp}^2(u)^{-1} \qquad (6)$$

$$N(u) = NPS_{SPD}(u)\,\gamma_{disp}^2\,MTF^2{}_{disp}(u) + NPS_{disp}(u) \qquad (7)$$

*Scene-and-Process-Dependent CPIQ metric*

The CPIQ metric is similarly revised, by substituting its input parameters for parameters that account for capture system and HVS scene-dependency. Keelan's [19] MF-IQM is defined by Eq. 8, where $QL_m$ is the overall quality loss, and $QL_i$ is the quality loss with respect to each attribute metric, $i$. $n_{max}$ is the power parameter, Eq. 9, where $QL_{max}$ is the maximum quality loss under the viewing conditions. The constants $c_1$ and $c_2$ are set to 2 and 16.9, respectively [21].

$$QL_m = \left(\sum_i (QL_i)^{n_{max}}\right)^{\left(\frac{1}{n_{max}}\right)} \tag{8}$$

$$n_{max} = 1 + c_1 \cdot \tanh\left(\frac{QL_{max}}{c_2}\right) \tag{9}$$

The IEEE P1858 CPIQ Standard's [12] texture loss and visual noise attribute metrics map the imaging chain acutance, $Q_T$, and the total visual noise metric, $\Omega$, to JND units, respectively, using curve-fitting functions derived from correlations with observers' data.

Equation 10 defines the imaging chain acutance [12], $Q_T$, where $MTF_{system}(u)$ is the capture system's direct dead leaves MTF [30]. $MTF_{disp}(u)$ is the display's modelled MTF. $CSF(u)$ is Johnson and Fairchild's luminance CSF [31], and $u$ and $u_{max}$ are the spatial frequency and cut-off frequency, respectively.

$$Q_T = \frac{\int_0^{u_{max}} MTF_{system}(u) \cdot MTF_{disp}(u) \cdot CSF(u)\, du}{\int_0^{\infty} CSF(u)\, du} \tag{10}$$

The total visual noise metric ($\Omega$) is computed by transforming captured uniform patch(es) through the linearized sRGB, CIEXYZ, and $AC_1C_2$ color spaces. The $AC_1C_2$ images are Fourier transformed and filtered with: i) Johnson and Fairchild's Luminance and Chrominance CSFs [31]. ii) the modelled display MTF [12]. iii) a high pass filter. After being inverse Fourier transformed, they are converted to the CIELAB color space via CIEXYZ. Ω is computed from the *L\*a\** covariance and the variances of *L\*, a\*,* and *b\*.*

We calculate the revised CPIQ metric in three stages: 1) the texture blur attribute metric is computed from a *Scene-and-Process-Dependent Acutance* measure, Eq. 10, where $MTF_{system}(u)$ is an appropriate SPD-MTF measure, $CSF(u)$ is one of the Barten CSF, cCSF, or cVPF, and all other parameters are as previously described. 2) the visual noise attribute metric is computed from a scene-and-process-dependent noise image, derived using the SPD-NPS framework. Either of the Barten CSF, cCSF or cVPF are used as the luminance CSF during spatial filtering. 3) Keelan's MF-IQM [19], Eq. 8, is used to compute the quality loss concerning #1 and #2.

**Novel Signal-to-Noise-Based Metrics**

In this section, we present the two novel metrics of this paper, the log NEQ and Visual log NEQ; they both employ the SPD-NEQ measure, which is also novel and is defined below.

The 2D NEQ, Eq. 11, is the standard measure of capture system signal-to-noise performance, versus spatial frequency $(u, v)$ [8]. $MTF(u, v)$ and $NPS(u, v)$ are the 2D MTF and NPS of the system, respectively. $\mu_A$ is the mean linear signal. Utilizing 1D MTFs and NPSs yields the 1D NEQ. The aforementioned limitations of current standard MTFs and NPSs are carried into the NEQ.

$$NEQ(u, v) = \frac{MTF^2(u,v)}{NPS(u,v)/\mu_A^2} \quad (11)$$

The SPD-NEQ is computed by substituting $MTF(u, v)$ and $NPS(u, v)$ with the 1D SPD-MTF and SPD-NPS, respectively.

*Log NEQ and Visual Log NEQ Metrics*

The log NEQ, Eq. 12, and Visual log NEQ, Eq. 13, model perceived image quality as the logarithm of the integral of the weighted SPD-NEQ measure, $NEQ_{SPD}(u)$. $MTF_{disp}^2(u)$ is the display's modeled

MTF [12], $u$ and $u_{max}$ are spatial frequency and cut-off frequency, respectively. $k_1$ and $k_2$ are calibration constants. $CSF(u)$ refers to one of the Barten CSF, cCSF, or cVPF. Both metrics apply minimal calibration compared to any other relevant metrics and relate directly to fundamental signal-to-noise relationships with perceived quality. Computing the logarithm of the NEQ is appropriate since it relates to Fechner's law [39], Shannon's Channel Capacity [13], and the PIC [17].

$$LogNEQ = k_1 \log_{10} \left( \int_0^{u_{max}} MTF_{disp}^2(u) NEQ_{SPD}(u) \frac{du}{u} \right) + k_2 \qquad (12)$$

$$LogNEQ_{Visual} = k_1 \log_{10} \left( \int_0^{u_{max}} CSF^2(u) MTF_{disp}^2(u) NEQ_{SPD}(u) \frac{du}{u} \right) + k_2 \qquad (13)$$

**Image Quality Metric Validation Methodology**

*Test Image Dataset*

Image quality evaluations were carried out for a series of test images, generated by linear and non-linear image capture pipelines, [9]. Test images were generated, starting with 14 original high-quality images, originating from a Canon DSLR camera equipped with professional lenses [40], [41]. These were downsized using bicubic interpolation and cropped to 512-by-512 pixels. Appendix A and Figure 2 show thumbnails of the original test scenes and their 1D power spectra, respectively, and were subsequently processed by two simulated image capture pipelines that were tuned to replicate real camera-phone artefacts at four exposure levels.

Both pipelines modelled the following processes identically: lens blur by convolution with a Gaussian model for a diffraction-limited lens' airy disk; linear SNRs of 10, 20, 40 and 80 at saturation by 2D Poisson noise; read and dark noise by Gaussian noise of higher standard deviation at lower SNRs; sensor quantum efficiency variations by scaling noise in the R, G and B channels by factors of 2, 1 and 3.3, respectively. Gain adjustments, noise floor removal, highlight recovery and Bayer color filter array (CFA) sampling were also kept constant for both pipelines.

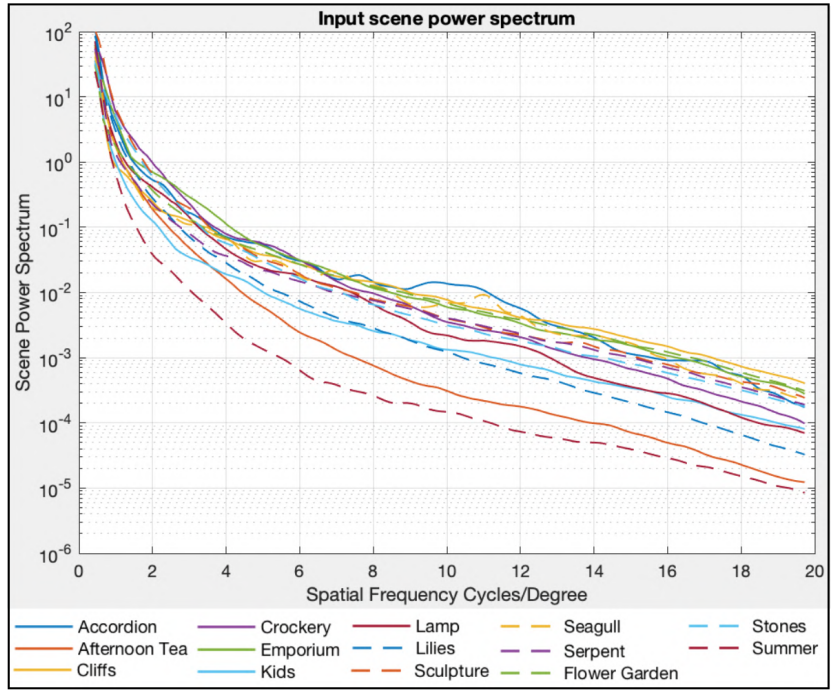

*Figure 2. Power spectra for the input scenes to the image capture simulations.*

Further, the linear pipeline employed the following linear ISPs: demosaicing by Malvar et al. [42], denoising by 2D Gaussian filtering and sharpening by the MATLAB™ *imsharpen* unsharp mask. The non-linear pipeline used the following non-linear content-aware ISPs: demosaicing by *One Step Alternating Projections (OSAP)* [43], denoising by *Block Matching and 3D Filtering (BM3D)* [44], sharpening of individual color channels by the *Guided Image Filter (GIF)* [45].

The denoising and sharpening filter input parameters, and the filters' opacities, Table 4, were tuned to optimize perceived output image quality, after combined sharpening and denoising at each SNR under the experimental conditions. Reducing the opacity, $P$, below 100%, lowered the intensity of filtering of the output image, $o(x,y)$, by blending the unfiltered, $g(x,y)$, and filtered images, $d(x,y)$ (Eq. 14). It was the only way to optimize subjectively the intensity of certain ISP filters at higher SNRs. It also tested the robustness of the IQMs employing SPD-MTFs and SPD-NPSs that are designed for filtered image signals and noise, respectively [9].

**Table 4. Optimal Denoising and Sharpening Filter Opacities.**

| Pipeline Type | ISP Type | Filter | Opacity | | | |
|---|---|---|---|---|---|---|
| | | | SNR 10 | SNR 20 | SNR 40 | SNR 80 |
| Linear | Denoising | Gaussian | 85% | 83% | 82% | 80% |
| | Sharpening | USM | 60% | 60% | 55% | 55% |
| Non-Linear | Denoising | BM3D | 87% | 86% | 86% | 85% |
| | Sharpening | GIF | 60% | 70% | 65% | 60% |

$$o(x,y) = \frac{P}{100} \cdot d(x,y) + \frac{100-P}{100} \cdot g(x,y) \qquad (14)$$

Fifty-six test images were output from each pipeline after the demosaicing, denoising and sharpening ISP stages. They covered all permutations of the 4 SNRs for all 14 input scenes. Test images from the linear pipeline were selected to represent both pipelines before denoising since the output images from both pipelines were nearly identical. A total of 280 test images were generated for evaluation.

*Psychophysical Image Quality Evaluations*

Psychophysical image quality ratings were recorded for each test image using the ISO 20462 Image Quality Ruler paradigm [3]. The image quality evaluations were carried out using a GUI interface and a ruler image set, generated by Allen [40], following ISO 20462 Part 3 [3]. The subjective quality of each test image was rated by selecting the corresponding ruler image, from a series of 30 ruler images of the same scene, which matched its quality. These ruler images differed in terms of sharpness only and ranged in quality from $SQS_2 = 3$ to $SQS_2 = 32$.

The experimental conditions, Figure 3, were similar to the requirements of ISO 20462 Part 3 [3] and the standard sRGB environment [46]. The EIZO ColorEdge CG245-W display was employed, with a pixel pitch of 0.27mm. It was calibrated to the sRGB color space with white point luminance of 120 cd/m$^2$. The Nyquist frequency was 20 cycles/degree at the 60cm viewing distance, which was restricted by a headrest. The ruler images were calibrated for these exact viewing conditions [40].

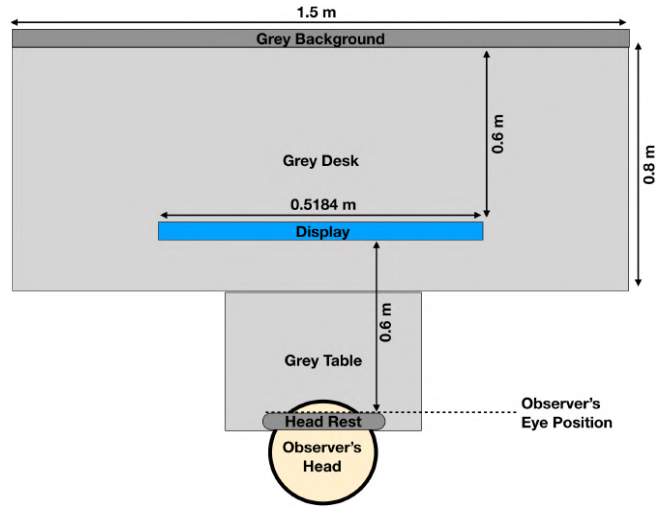

*Figure 3. Layout of the laboratory equipment for the image quality evaluations.*

Twenty-seven observers participated, including 10 females and 17 males of various ethnicities, with approximate age range of 20 to 55 years old. Six had prior experience of comparable image quality evaluations. Observers wore corrective spectacles/lenses if required for the viewing distance. Their visual acuity was confirmed for the given viewing conditions using a Snellen near vision test card [47].

*Metric Variant Computation and Calibration*

Variants of the revised and proposed IQMs were generated, and calibrated, to be benchmarked against the observer image quality ratings. A total of 332 variants were created, employing different permutations of the MTF, NPS and CSF parameters in Tables 1, 2 and 3, respectively. The sensitivity of these variants to imaging system and visual scene-dependency varied considerably.

SPD-MTFs and SPD-NPSs were computed using ten replicates by adapting MATLAB[TM] code for Burns' direct dead leaves MTF implementation [22]. The CPIQ visual noise metric was computed using Baxter and Murray's [48] MATLAB[TM] implementation [49].

Variants of the SQRIn, PIC, log NEQ and Visual log NEQ were calibrated to the SQS$_2$ scale by: i) setting $k_2$ to zero as [27]. ii) optimizing $k_1$, so the mean of the output scores at SNR 80 (best quality) for all test images before denoising equaled the mean of the respective observer ratings. When employed, the cCSF/cVPF were normalized to the same integrated area as the Barten CSF. The metric scores were thus affected by scene-dependent changes in the cCSFs/cVPFs shape, but not changes in their magnitude.

Variants of the CPIQ metric were calibrated to $SQS_2$ units by subtracting the predicted quality loss (i.e. $QL_m$ in Eq. 8) from the $SQS_2$ value of 23 corresponding to the $SQS_2$ of the input scenes to the simulations. The cCSF, cVPF and Barten CSF, when employed, were normalized to the same integrated area as Johnson and Fairchild's Luminance CSF, as used by the IEEE P1858 CPIQ Standard [12].

## Results

### *Results from Quality Evaluations*

Figure 4 presents the observer ratings from the image quality evaluations. They were not calibrated according to the average scene relationship [3] since Allen [40] found this action removed virtually all scene-dependency from the data. Such scene-dependencies demanded consideration in this paper, since the purpose of the study is to test whether capture system and HVS scene-dependency can be accounted for successfully in the selected metrics. Uncalibrated ratings were used successfully when evaluating predictions of non-linear JPEG and JPEG 2000 compressed image quality by IQMs [40].

The ratings were generally higher after combined denoising and sharpening, for which the tuning was optimized. As expected, non-linear ISPs caused the greatest scene-dependent variations. They also produced higher quality images than the respective linear ISPs at lower SNRs, because they preserved genuine image signal content or mitigated the amplification of noise. Variances of the scenes' susceptibility and observers' sensitivity to quality losses increased at lower SNRs. The former was lower than the latter, as observed previously by Keelan [19].

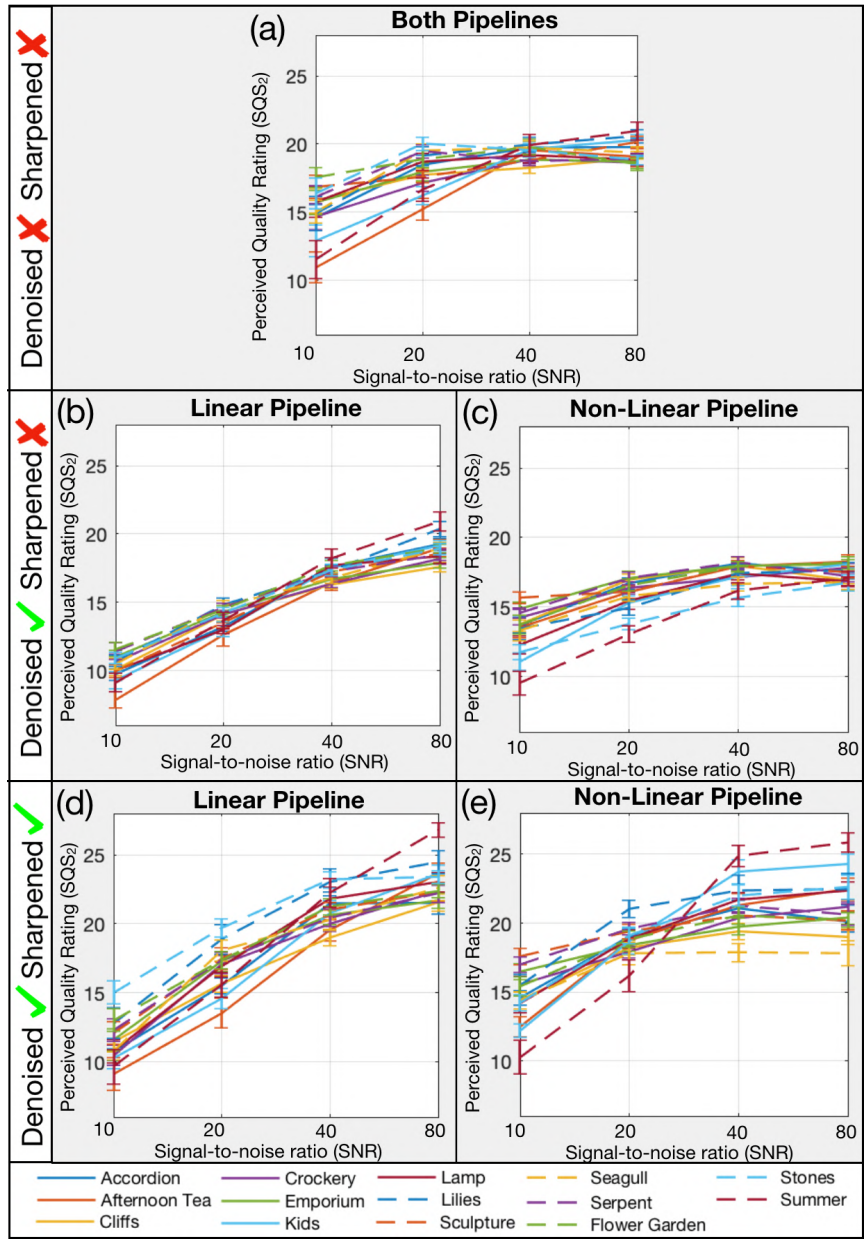

*Figure 4. Observer image quality ratings (SQS$_2$) for each scene after different stages of linear and non-linear processing. Error bars show standard error.*

***Benchmarking of Image Quality Metrics***

Table 5 benchmarks the most and least accurate variants of each metric, in terms of their Mean Absolute Error (MAE). The MAE describes specifically the mean difference in SQS$_2$ units between the metric scores and the ideal linear relationship with the observer quality ratings, shown by the pink line in Figures 6-8. Thus, if a variant has a MAE of 2 then it can be expected to predict the perceived quality of a given image with an accuracy of ±2 JNDs.

**Table 5. Benchmarking of Highest and Lowest Performing Variants of Each Metric, in Terms of Mean Absolute Error (MAE)**

**Highest Accuracy Variants of Each Metric**

| Metric Name | Input Parameters | | | Pipeline Type | MAE |
|---|---|---|---|---|---|
| | Noise Measure | MTF | CSF | | |
| CPIQ | Uses Pictorial Image Replicates | Pictorial Image SPD-MTF | Barten | Non-Linear | 1.82 |
| | | | | Linear | 1.57 |
| Visual Log NEQ | Mean Pictorial Image SPD-NPS | Dead Leaves SPD-MTF | Barten | Non-Linear | 1.68 |
| | | | | Linear | 2.25 |
| Log NEQ | Pictorial Image SPD-NPS | Dead Leaves SPD-MTF | N/A | Non-Linear | 2.16 |
| | | | | Linear | 2.04 |
| PIC | Pictorial Image SPD-NPS | Dead Leaves SPD-MTF | Any of Barten, cCSF or cVPF | Non-Linear | 2.50 |
| | | | | Linear | 2.47 |
| SQRIn | Pictorial Image SPD-NPS | Pictorial Image SPD-MTF | Any of Barten, cCSF or cVPF | Non-Linear | 5.46 |
| | | | | Linear | 4.92 |

**Lowest Accuracy Variants of Each Metric**

| Metric Name | Input Parameters | | | Pipeline Type | MAE |
|---|---|---|---|---|---|
| | Noise Measure | MTF | CSF | | |
| CPIQ | Uses Dead Leaves Replicates | Direct Dead Leaves MTF | cVPF | Non-Linear | 9.88 |
| | | | | Linear | 8.82 |
| Visual Log NEQ | Uniform Patch NPS | Pictorial Image SPD-MTF | cVPF | Non-Linear | 2.48 |
| | | | | Linear | 2.38 |
| Log NEQ | Uniform Patch NPS | Pictorial Image SPD-MTF | N/A | Non-Linear | 3.00 |
| | | | | Linear | 2.13 |
| PIC | Uniform Patch NPS | Direct Dead Leaves MTF | Any of Barten, cCSF or cVPF | Non-Linear | 3.15 |
| | | | | Linear | 2.53 |
| SQRIn | Uniform Patch NPS | Direct Dead Leaves MTF | Any of Barten, cCSF or cVPF | Non-Linear | 6.94 |
| | | | | Linear | 5.34 |

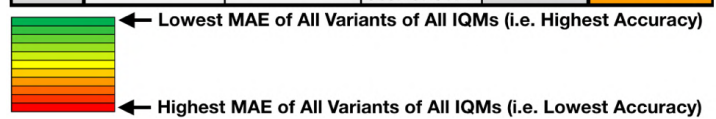
← Lowest MAE of All Variants of All IQMs (i.e. Highest Accuracy)

← Highest MAE of All Variants of All IQMs (i.e. Lowest Accuracy)

When evaluating the robustness of each metric to changes in its input parameters, the main factor we consider is the accuracy of its most accurate variant. The range of accuracy across all variants of the metric is a further important factor we discuss, Figure 5.

The CPIQ metric produced the variants with the highest overall accuracy, followed closely by the new Visual log NEQ and log NEQ metrics of this paper, and the PIC. However, the CPIQ metric was highly sensitive to changes in its input parameters, Figure 5, in particular the CSF. This led to it also producing the least accurate variants of all. We expect this unpredictable behavior to be due to curve-fitting "forcing" the metric to its original input parameters.

The STV-IQMs – which include the PIC, SQRIn, Log NEQ and Visual Log NEQ – showed greater consistency when their input parameters were changed, especially for the linear pipeline, Figure 5. For the non-linear pipeline, the Visual Log NEQ produced the most accurate variants out of all the metrics tested, demonstrating the power of these simpler signal-to-noise-based metrics.

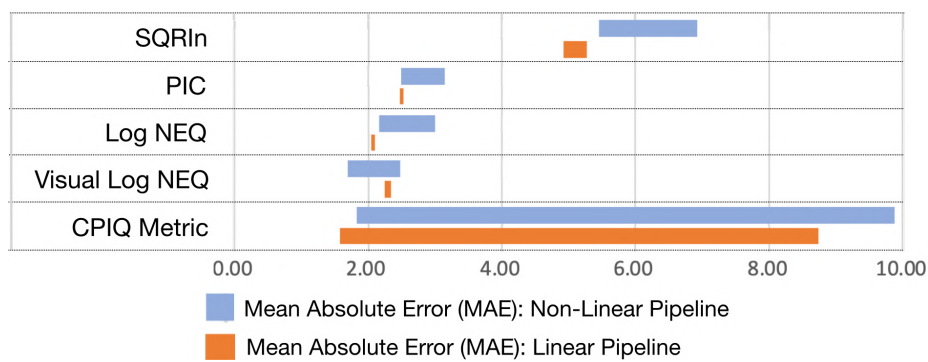

*Figure 5. The range of Mean Absolute Error (MAE) values across all variants of each metric.*

Employing the various SPD-NPSs and SPD-MTFs generally improved the metrics' accuracy for the non-linear pipeline. This was despite the reduced ISP filter opacities giving the current standard MTF/NPS measures an advantage. Tables benchmarking the MAE of every variant of each IQM demonstrated the consistency of these improvements. IQMs utilizing the SPD-NPSs and SPD-MTFs were often of comparable accuracy for the linear and non-linear pipelines.

The most accurate variants of each metric used SPD-NPSs or noise images derived from scenes. This demonstrates the robustness of the SPD-NPS framework and the benefits of accounting most comprehensively for system noise scene-dependency. In contrast, the uniform patch NPS was used by the vast majority of the lowest accuracy variants, substantiating previous observations that the dead leaves SPD-NPS is the more generally appropriate measure [9].

When considering the appropriateness of the revised MTF, NPS and CSF parameters, evaluating their effect upon the accuracy of the log NEQ and Visual log NEQ is of particular relevance, due to the simplicity and "purity" of these new metrics. The most accurate variants of these IQMs used the dead leaves SPD-MTF. This suggests that the negative effect of bias in the pictorial image SPD-MTF outweighed the benefit of accounting more comprehensively for system scene-dependency. Nevertheless, the CPIQ metric and SQRIn performed most accurately with pictorial image SPD-MTFs. This demonstrates the potential benefits of the measure, which we suggest should be developed further to address measurement bias, especially when considering the success of the respective SPD-NPS measure.

The highest performing variants of each metric all used the Barten CSF. Implementing the cCSF or cVPF generally affected negatively the metrics' accuracy. This was also the case when the cCSF/cVPF were not normalized. Changing the CSF parameter did not affect the accuracy of the PIC or SQRIn significantly. This was because the high display luminance reduced visual noise to a level where changes in the CSF parameter cancelled themselves out.

Benchmarking with respect to the Root Mean Square Error (RMSE) and Spearman's Rank Order Correlation Coefficient (SROCC) displayed comparable trends to those in Table 5.

*Further Analysis of Metric Correlations*

We analyze in this section correlations between output scores of the selected IQM variants and the observer quality ratings. The analysis demonstrates the typical IQM behavior and any significant changes resulting from revision of their input parameters.

We observe that, when the log NEQ employs the more relevant dead leaves SPD-NPS (Figure 6b) instead of the uniform patch NPS (Figure 6a), it no longer overestimates perceived image quality after non-linear sharpening or denoising; interestingly, it also produces similar correlations to the linear pipeline. Utilizing SPD-NPSs derived from pictorial scenes led to further improvements, Figure 6c. The above was also true for the PIC, SQRIn and Visual log NEQ. Respective variants of the latter metric were similar to those in Figure 6, showing improved accuracy.

The SQRIn overestimated perceived image quality at high SNRs, resulting in a curved distribution, Figure 7. This corroborates Töpfer and Jacobson's observation [17] that under some conditions, the metric does not describe perceived image quality linearly with noise, and in accurate JND units. This is expected to be due to the underestimation of the perceived intensity of the noise at near-threshold levels [17]. The PIC displayed the same behavior to a lesser extent. We conclude that our simulations exceeded the SNR threshold to which the SQRIn and PIC apply, which is understandable since both metrics were developed and calibrated for analog systems with generally higher noise.

The CPIQ metric (Figure 8) was linearly correlated with the observer ratings for all permutations of the input parameters. These correlations did not follow the 'ideal' linear relationship with gain = 1 and offset = 0, however. The main cause of this gain in the original CPIQ metric's correlation, Figure 8a, was the effects of content-aware ISPs, although its correlation still suffered from some offset for the linear pipeline. Employing the Barten CSF and the various SPD-MTF and SPD-NPS measures reduced generally both the gains and offsets of the linear regression and rendered the correlations for the linear and non-linear pipelines more similar. This was despite the limitations of changing the input parameters in a metric that employs pre-calibration with observers' ratings.

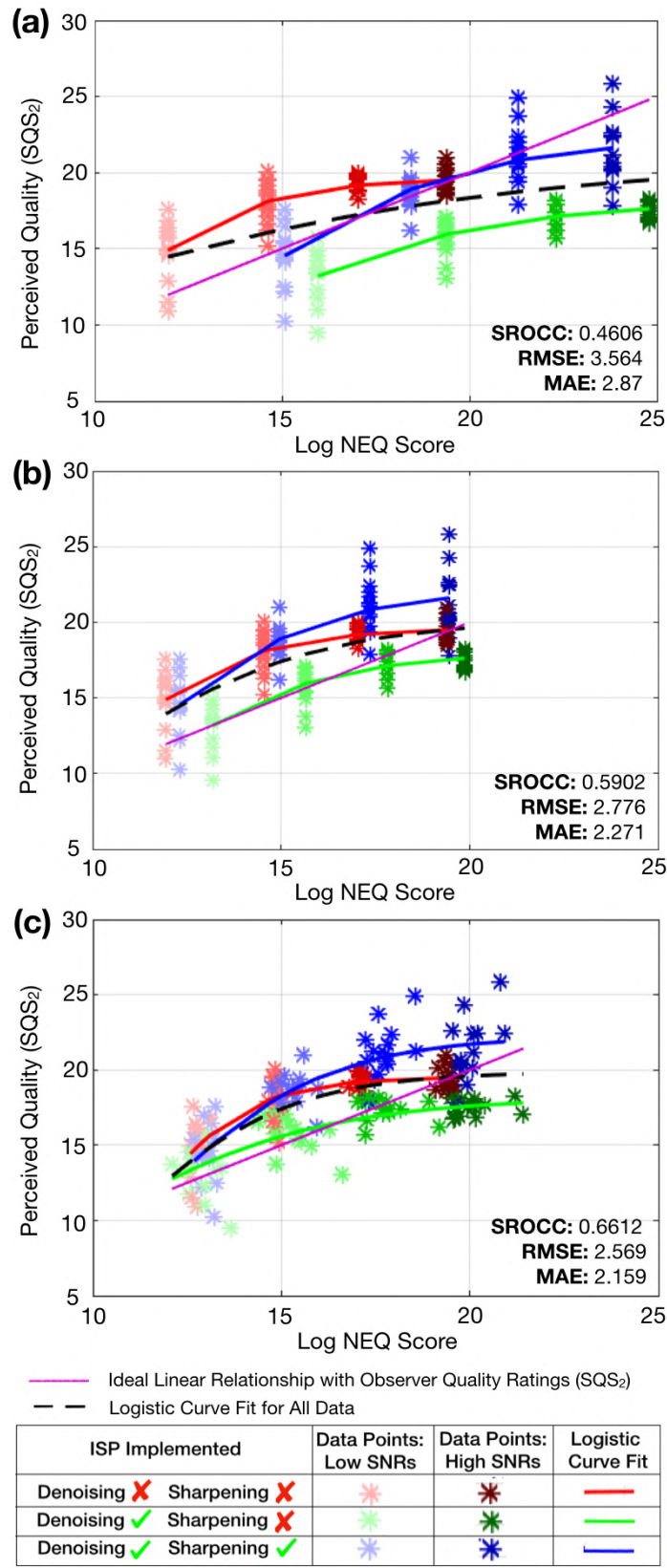

*Figure 6. Observer image quality ratings versus output scores of log NEQ variants employing: a) the direct dead leaves MTF [30] and uniform patch NPS; b) the direct dead leaves MTF [30] and dead leaves SPD-NPS; c) the pictorial image SPD-NPS and dead leaves SPD-MTF. (c) is the most accurate log NEQ variant. All test images were generated by the non-linear pipeline.*

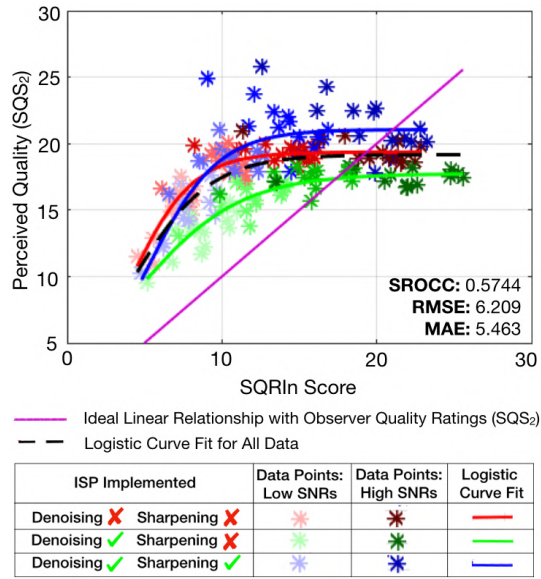

*Figure 7. Observer image quality ratings versus metric scores for the highest performing SQRIn variant, which employed the Pictorial Image SPD-MTF and SPD-NPS. All test images were generated by the non-linear pipeline.*

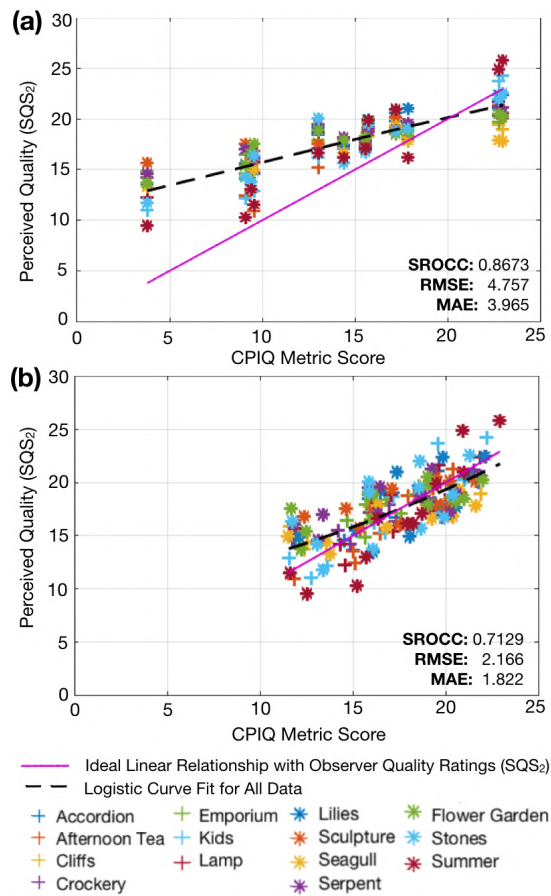

Figure 8. Observer image quality ratings versus output scores of CPIQ metric variants employing: (a) input parameters specified in the IEEE P1858 CPIQ standard *[12]* including the direct dead leaves MTF *[30]*, uniform patch noise image and Johnson and Fairchild luminance CSF *[31]*; (b) the pictorial image SPD-MTF, noise images computed from pictorial image replicates and the Barten *[32]* CSF. (b) is the highest performing CPIQ metric variant. All test images were generated by the non-linear pipeline.

**Conclusions**

Two novel metrics – the log NEQ and Visual log NEQ – were presented in this paper. Leading engineering IQMs, modeling spatial image quality, were also revised, including the IEEE P1868 CPIQ metric [12], SQRIn [16] and PIC [17]. Substituting the MTF, NPS and CSF parameters with equivalent scene-dependent measures (i.e. SPD-MTF, SPD-NPS, cCSF and cVPF) created variants of each metric. All metric variants were benchmarked, using different permutations of these measures.

The log NEQ and Visual log NEQ relate directly to the novel SPD-NEQ signal-to-noise measure and apply minimal calibration, which is well prescribed and depends on the given experimental conditions only. Thus, revising their input parameters does not, in theory, offer any violation to these metrics. Variants of these IQMs were generally more accurate than those of the comparable SQRIn and PIC metrics, and almost as accurate as those of the CPIQ metric. This not only demonstrates the relevance of the fundamental NEQ and SPD-NEQ measures to quality modeling, but it is also a very exciting outcome, considering the metrics' simplicity and 'purity'.

The CPIQ metric produced the most accurate variants of all, but the other variants ranged widely in terms of accuracy. We expect the latter was due to the CPIQ metric's pre-calibration and suggest the accuracy of the log NEQ and Visual log NEQ variants informs better regarding the appropriateness of the various input parameters.

Implementing the SPD-NPSs consistently improved the accuracy of all IQMs. This was particularly the case for SPD-NPSs derived from scenes. This indicates the SPD-NPS framework is more suitable for quality modelling than the uniform patch NPS and corroborates previous conclusions that it is robust, and accounts for scene-dependencies in temporally varying system noise [9].

For three out of five metrics tested, the most accurate variants employed the dead leaves SPD-MTF. The other two metrics performed most accurately using the pictorial image SPD-MTF. Thus, there appear to be trade-offs between the negative effect of bias in the latter measure, and the positive effect of it accounting more comprehensively for system signal transfer scene-dependency.

Research efforts in deriving SPD-MTFs from extracted pictorial scene edges continues [50]. It remains to be seen whether the resultant MTFs are more suitable parameters for image quality modelling.

Utilizing contextual detection, cCSF, or discrimination models, cVPF, that account for visual masking did not improve IQM accuracy, regardless of whether or not they were normalized.

We recommend further investigations to establish whether: i) contextual visual models have a genuine role in quality modelling. ii) validating the IQMs using real capture systems corroborates the results of this paper. iii) the IQMs of this paper describe accurately the quality of compressed images (JPEG and JPEG 2000).

The novel and revised IQMs of this paper, and their SPD-MTF, SPD-NPS, cCSF and cVPF input parameters, represent a new paradigm of image quality models, imaging systems characterization measures, and human visual models that account for relevant scene-dependencies in spatial imaging system and visual performance. The benefits of these more complex measures/metrics are demonstrated in this paper, and in our previous work [51].

## Appendix A: Input Scenes to the Simulations

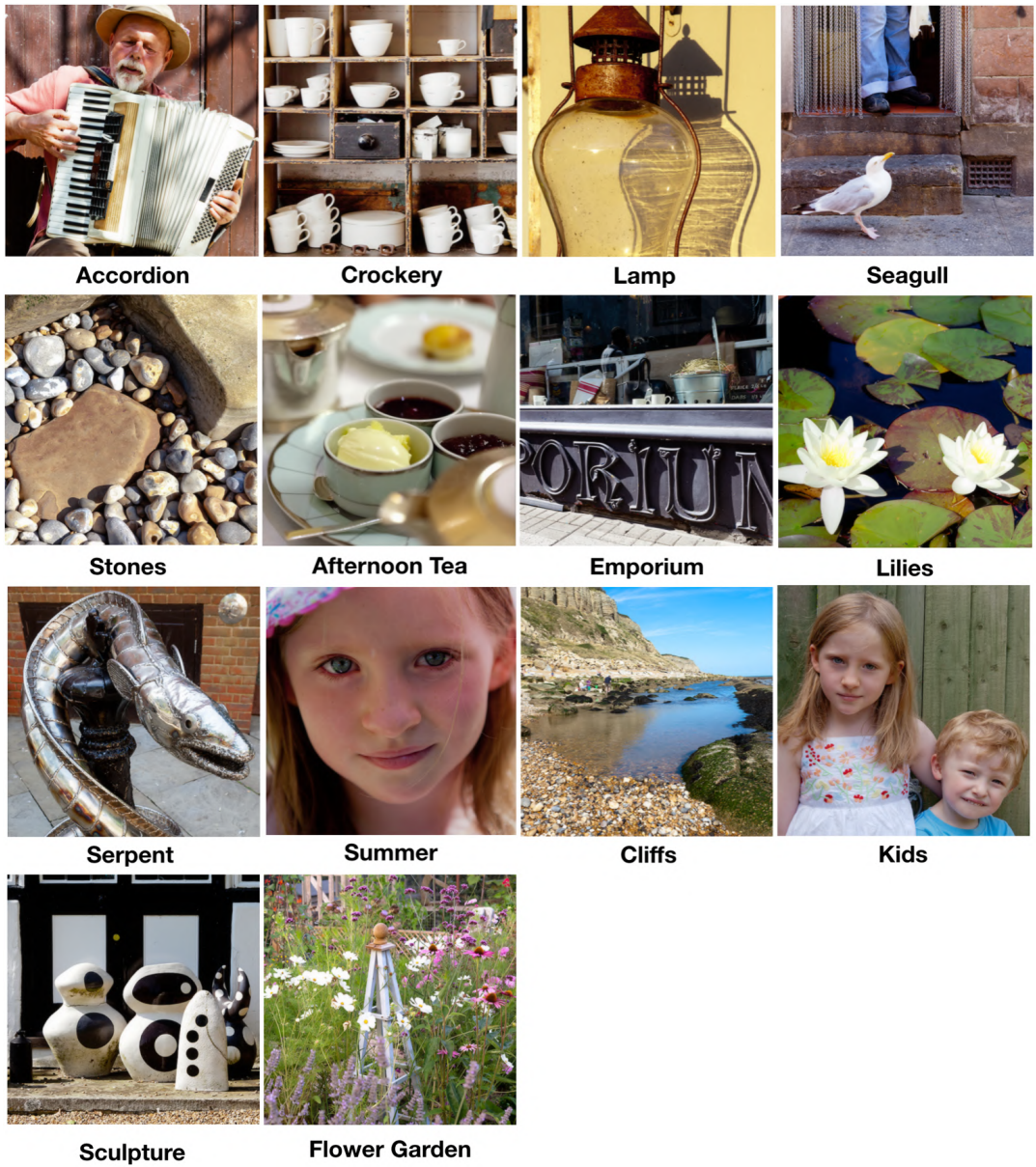

Figure 9. Input scenes to the image capture simulations. All scenes were captured and processed by Allen *[40]*.

## Abbreviations

cCSF…………………………………………………………………………………..Contextual Contrast Sensitivity Function

cVPF………………………………………………………………………………………Contextual Visual Perception Function

CPIQ Metric……………………………………………………………………………. IEEE P1858 CPIQ Standard Metric

Log NEQ……………………………………………………………………………………… Log Noise Equivalent Quanta Metric

MF-IQM……………………………………………………………………………….. Multivariate Formalism Image Quality Metric

PIC……………………………………………………………………………….. Perceived Information Capacity Metric

SPD-MTF………………………………………………………………………… Scene-and-Process-Dependent MTF

SPD-NEQ………………………………………………………………...…… Scene-and-Process-Dependent NEQ

SPD-NPS……………………………………………………………………….. Scene-and-Process-Dependent NPS

SQRIn……………………………………………………………………….. Square Root Integral with Noise Metric

STV-IQM………………………………………………………………….. Signal Transfer Visual Image Quality Metric

Visual log NEQ……...…………………………………………….. Visual Log Noise Equivalent Quanta Metric

**Author Biography**

Edward Fry is a PhD. student at the University of Westminster, UK, in the field of imaging systems performance measurement and image quality modeling. His research focuses on scene-dependency in imaging systems performance, and its impact upon overall perceived image quality.